\documentclass[english]{cbctq}

\usepackage[english]{babel}
\usepackage{amsmath,amssymb,amsfonts}
\usepackage{graphicx}
\usepackage{cite}
\usepackage{booktabs}
\usepackage{array}
\usepackage{url}
\usepackage{microtype}
\usepackage{tikz}

\newtheorem{proposition}{Proposition}
\newtheorem{definition}{Definition}

\begin{document}

\title{Operational Roles of QRNG-Derived Quantum Entropy in Bitcoin Proof-of-Work Architectures}

\author{
Ricardo Fernandes da Silva and Paulo Vitor Batista Santos%
\thanks{
Ricardo Fernandes da Silva, Academic Department of Physics, Federal Technological University of Paraná, Curitiba, Brazil, e-mail: ricardosilva@utfpr.edu.br;
Paulo Vitor Batista Santos, Foton Institute of Quantum Sciences and Technologies, Porto Alegre, Brazil, e-mail: eng.computacao\underline{ }paulovitor@hotmail.com. 
}
}

\maketitle

\markboth{Congresso Brasileiro de Ci\^encias e Tecnologias Qu\^anticas (CBCTQ 2026) -- September 21--25, 2026, Innovation District of Cantareira, Niter\'oi--RJ, Brazil}{}

\begin{abstract}
Replacing classical entropy with QRNG output does not change honest Bitcoin PoW success probability when candidate headers remain distinct. The original contribution of this paper is a reproducible benchmark that locates and measures the operational value of quantum entropy in hybrid quantum-classical mining infrastructure through two scheduler-level observables, the entropy-efficiency factor $\eta$ and the reboot-diversity index $\rho$. Monte Carlo and scheduler simulations with confidence intervals show parity for competent deterministic and strong-classical baselines, while QRNG value emerges in assurance-oriented scenarios involving correlated restart faults, namespace reuse, and entropy provenance. The study is therefore positioned as a simulation-based validation framework rather than as a device-level QRNG demonstration; hardware-in-the-loop validation with recorded or live QRNG streams is identified as the next experimental step.
\end{abstract}


\section{Introduction}
Quantum random number generators (QRNGs) extract entropy from genuinely quantum processes and are increasingly relevant to cryptography, certified randomness, and security-critical platforms \cite{herrero,ma,pironio,acin,mannalath}. Bitcoin, in turn, remains the best-known Proof-of-Work (PoW) blockchain, where miners repeatedly evaluate double-SHA-256 on candidate headers until the digest falls below the network target \cite{nakamoto,btcdev,antonopoulos,shs}. Because mining is often described informally as a random search, QRNGs are sometimes framed as possible mining accelerators.

This paper defends a narrower and technically stronger claim. For honest mining over distinct candidate headers, block discovery is governed by the number of effective trials, not by whether header ordering is deterministic, classically pseudorandom, or QRNG-assisted. The scientifically relevant role of quantum entropy therefore lies in the surrounding control plane: entropy roots, namespace freshness, restart robustness, and auditable provisioning.
Figure~\ref{fig:entropy_mapping} summarizes the equivalence in effective sampling induced by different entropy strategies under distinct-header testing. When QRNG is used, it acts at the entropy root of the control plane, supporting scheduler state, namespace assignment, and reseeding policy, while the PoW hashing engine itself remains unchanged. Accordingly, any QRNG-related effect arises from orchestration quality and restart behavior rather than from a modification of the per-hash success law.
% no preâmbulo
\usetikzlibrary{positioning,arrows.meta}

% no corpo do texto
\begin{figure}[t]
\centering
\begin{tikzpicture}[
    >=Latex,
    node distance=5mm and 5mm,
    every node/.style={
        draw,
        rectangle,
        rounded corners=2pt,
        align=center,
        inner sep=3pt,
        font=\footnotesize,
        minimum height=8mm
    }
]

\node[text width=2.2cm] (det) {Deterministic\\Enumeration};
\node[text width=2.2cm, right=of det] (csprng) {CSPRNG\\(Classical)};
\node[text width=2.2cm, right=of csprng] (qrng) {QRNG\\Assisted};

\node[draw=none, below=9mm of csprng, font=\footnotesize] (space) {$\rightarrow$ Uniform Hash Space};

\draw[->, line width=0.4pt] (det.south) -- (space.north west);
\draw[->, line width=0.4pt] (csprng.south) -- (space.north);
\draw[->, line width=0.4pt] (qrng.south) -- (space.north east);

\end{tikzpicture}
\caption{Different entropy strategies converge to the same effective sampling distribution over the hash space. Under distinct-header testing, deterministic enumeration, strong classical pseudorandomization, and QRNG-assisted scheduling differ in entropy provisioning, but not in the first-order PoW success law.}
\label{fig:entropy_mapping}
\end{figure}
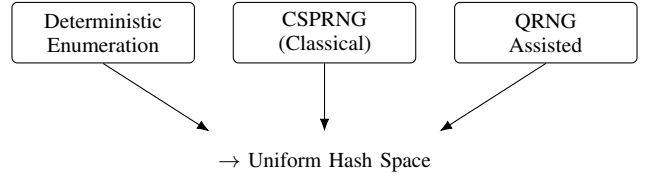
The contribution is not a new PoW law, but a \emph{reproducible operational benchmark} for the disciplined integration of a quantum primitive into hybrid quantum-classical infrastructure. Specifically, the paper contributes: i) a fair comparison among deterministic, strong-classical, and QRNG-assisted schedulers under identical PoW logic; ii) two measurable observables, the entropy-efficiency factor $\eta$ and the reboot-diversity index $\rho$, both recoverable from controller or pool logs; and iii) a correlated-restart protocol with sensitivity analysis that identifies the operational regimes in which QRNG adoption is justified on assurance grounds. Accordingly, the relevant technical question is not whether QRNG changes the PoW success law, but how quantum entropy should be integrated, measured, and audited in a larger classical distributed system.

This paper does not report measurements from a physical QRNG device or from quantum hardware. Instead, it defines a reproducible benchmark and a set of observables intended to guide future device-in-the-loop experiments. This scope is important because the expected contribution of QRNG in Bitcoin-related infrastructure is not acceleration of the hashing process, but improved assurance, provenance, and robustness of entropy-dependent control services.

\section{Related Work and Technical Positioning}

The QRNG literature has firmly established the relevance of physical unpredictability, entropy estimation, and deployment models ranging from laboratory-scale devices to integrated photonic implementations \cite{herrero,ma,mannalath,marangon,crampton}. More recently, the focus has expanded from isolated QRNG characterization toward system-level integration, including entropy-as-a-service architectures, cloud-based provisioning, and QRNG-backed post-quantum communication stacks \cite{vassilev,huang,blanco2025,blanco2026}. This body of work increasingly treats quantum randomness as an operational resource that can be embedded into hybrid infrastructures rather than as a purely standalone physical capability.

At the same time, the Bitcoin and PoW literature has extensively characterized mining mechanics, pool coordination, adversarial incentives, and operational attack surfaces \cite{nakamoto,bonneau,gervais,rosenfeld,kroll,eyal,minersdilemma,luu}. However, the interface between these two literatures remains insufficiently developed. In blockchain-oriented discussions, QRNG is often assessed in ways that blur the distinction between consensus probability and entropy assurance. When explicit comparisons are made, QRNG is frequently contrasted with weak or poorly specified classical sources, rather than with a properly engineered deterministic random bit generator (DRBG) consistent with NIST guidance \cite{sp90a,sp90b,sp90c}. Moreover, few studies introduce observables that are directly recoverable from operational logs and therefore suitable for reproducible infrastructure-level evaluation.

The stance adopted in this paper is deliberately narrow and therefore methodologically stronger. We do not claim any protocol-level quantum advantage in Bitcoin mining. Rather, we develop a fair benchmarking framework and an instrumentation-oriented vocabulary for assessing when QRNG insertion is technically justified in mining-adjacent infrastructure. The central novelty is thus methodological: a disciplined procedure for evaluating the role of a quantum primitive within hybrid quantum-classical systems.

\section{Analytical Framework and Measurable Observables}
A Bitcoin block is accepted when
\begin{equation}
H(H(B_h)) < T,
\end{equation}
where $H$ is SHA-256, $B_h$ is the 80-byte block header, and $T$ is the target implied by network difficulty \cite{nakamoto,antonopoulos}. Under the standard assumption that double-SHA-256 outputs on distinct inputs are computationally indistinguishable from uniform values, one distinct tested header succeeds with probability
\begin{equation}
p = \frac{T}{2^{256}}.
\end{equation}

\begin{proposition}[Consensus neutrality]
If a miner tests $k$ distinct candidate headers, then the probability of finding at least one valid block is
\begin{equation}
P(1)=1-(1-p)^k \approx kp \qquad (p \ll 1),
\end{equation}
independently of whether the ordering of those $k$ headers is generated by deterministic enumeration, a strong classical DRBG, or a QRNG-assisted entropy root.
\end{proposition}

The proposition is a baseline, not the novelty of the paper. It only rules out a direct protocol-level mining advantage from QRNG substitution alone. The operational question is whether imperfect entropy handling reduces the number of \emph{distinct} work opportunities actually explored. Let $k$ be the number of issued work units in a measurement window and let $u \leq k$ be the number of distinct work units effectively explored.

\begin{definition}[Entropy-efficiency factor]
\begin{equation}
\eta = \frac{u}{k}, \qquad 0<\eta\leq 1.
\end{equation}
\end{definition}

To make $\eta$ measurable from scheduler logs, we use the work-unit identifier
\begin{equation}
w=(h_{\mathrm{prev}},m,t_b,n_b,x_b),
\end{equation}
where $h_{\mathrm{prev}}$ is the previous block hash, $m$ is the Merkle-root state, $t_b$ is a timestamp bucket, $n_b$ is a nonce range, and $x_b$ is an extranonce or equivalent namespace slice. Then $u$ is estimated from the number of unique $w$ values observed in controller or pool logs. The first-order discovery law becomes
\begin{equation}
P(\eta)=1-(1-p)^{\eta k}\approx \eta kp.
\end{equation}

Throughput parity does not imply equal assurance. A deployment can keep $\eta \approx 1$ in steady state and still be fragile under synchronized reboots or image-based recovery. We therefore define a second observable.

\begin{definition}[Reboot-diversity index]
For $W$ workers and reboot event $t$,
\begin{equation}
\rho = \frac{1}{W}\sum_{i=1}^{W} \mathbf{1}\!\left[\sigma_i^{(t)} \neq \sigma_i^{(t-1)}\right],
\end{equation}
where $\sigma_i^{(t)}$ is the namespace seed or domain-separation token assigned to worker $i$ immediately after reboot $t$.
\end{definition}

Here $\rho=1$ means complete post-boot diversity, whereas smaller values indicate repeated state. Unlike $\eta$, $\rho$ is not a consensus quantity; it is an infrastructure-assurance metric. Together, $\eta$ and $\rho$ locate where entropy quality can matter without altering PoW itself. A complementary finite-namespace approximation is also useful: if $k$ work units are drawn with replacement from a namespace of size $N$, then
\begin{equation}
\mathbb{E}[\eta] \approx 1-\frac{k-1}{2N} \qquad (k\ll N),
\end{equation}
which explains how overlap can arise from narrow or reused namespaces even without adversarial manipulation.

\section{Simulation Protocol and Validation Metrics}
All baselines share the same header-testing logic, the same work-unit format, the same nominal work volume, and the same reseed cadence; only the entropy/control policy changes. One simulation episode is defined as a reboot event followed by issuance of 256 work units to each of 64 workers. For each issued unit, the simulator records the tuple $w$ above and the post-boot digest of the worker namespace token. This makes the protocol reproducible from ordinary controller logs rather than unrealistic ASIC-level traces. No physical QRNG stream is used in the present implementation; the QRNG-assisted condition is modeled as an idealized high-entropy root to isolate the operational effect of entropy provisioning.

\begin{table}[t]
\caption{Reproducible protocol parameters.}
\centering
\footnotesize
\begin{tabular}{>{\raggedright\arraybackslash}p{0.41\linewidth} p{0.49\linewidth}}
\toprule
Parameter & Value \\
\midrule
PoW validation sweep & $k=10^3$, $p=10^{-3}$, $3\times 10^4$ windows per $\eta$ \\
Scheduler episodes & $10^4$ reboot episodes \\
Workers per episode & $W=64$ \\
Issued work per worker & 256 units \\
Correlated-restart control & 20\% workers restored from 6 cached images \\
State reuse in degraded control & 50\% of restored workers keep previous local state before reseeding resumes \\
Confidence interval & 95\% CI on episode means \\
\bottomrule
\end{tabular}
\label{tab:protocol}
\end{table}
Figure~\ref{fig:architecture} summarizes the system-level separation adopted in the simulations. Across all baselines, the PoW hashing path, work-unit structure, and nominal workload remain fixed, while only the entropy provisioning and scheduler-control layer are varied. This separation is central to the simulation design because it isolates QRNG-related effects to orchestration, namespace assignment, and reseeding policy rather than to the hashing engine itself.
\usetikzlibrary{positioning,arrows.meta}

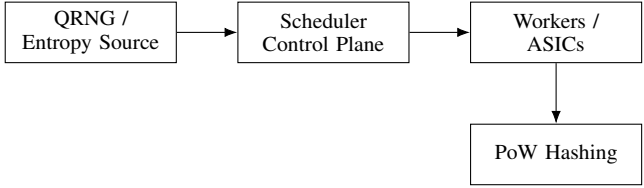
\begin{figure}[t]
\centering
\begin{tikzpicture}[
    >=Latex,
    node distance=6mm and 8mm,
    block/.style={
        draw,
        rectangle,
        align=center,
        minimum height=8mm,
        text width=2.05cm,
        inner sep=3pt,
        font=\footnotesize
    }
]

\node[block] (qrng) {QRNG /\\Entropy Source};
\node[block, right=of qrng] (scheduler) {Scheduler\\Control Plane};
\node[block, right=of scheduler] (miners) {Workers /\\ASICs};
\node[block, below=8mm of miners] (pow) {PoW Hashing};

\draw[->, line width=0.4pt] (qrng) -- (scheduler);
\draw[->, line width=0.4pt] (scheduler) -- (miners);
\draw[->, line width=0.4pt] (miners) -- (pow);

\end{tikzpicture}
\caption{Architectural separation between entropy provisioning, scheduler control, and PoW execution. QRNG, when present, affects seeding, reseeding, and namespace management in the control plane, while the PoW hashing engine remains unchanged.}
\label{fig:architecture}
\end{figure}
The four baselines are as follows. \textbf{B1} uses deterministic partitioning with disjoint counter spaces and explicit reboot counters for domain separation. \textbf{B2} is a strong classical baseline in which namespace tokens are derived from an independent classical entropy source plus a properly conditioned DRBG chain consistent with the NIST SP 800-90 series \cite{sp90a,sp90b,sp90c}. This strong-classical baseline is methodologically important because it prevents an unfair comparison between QRNG and a deliberately weak or underspecified classical entropy source. In practical terms, \textbf{B2} represents the best classical engineering case considered in this study: an independent classical entropy source combined with a properly instantiated and conditioned DRBG chain under standard NIST assumptions. Under this benchmark, parity between \textbf{B2} and \textbf{B3} in fault-free regimes should not be read as a null result, but as evidence that QRNG is properly evaluated on assurance, provenance, and restart robustness rather than on a fictitious modification of the PoW success law. \textbf{B3} is QRNG-assisted in the modeled sense: the scheduler logic and DRBG chain are preserved, but the entropy root is replaced by an idealized high-entropy seed stream representing the interface that would be supplied by a physical QRNG in a device-in-the-loop implementation. \textbf{B4} is a correlated-restart control: after each reboot, 20\% of workers are restored from one of six cached images, and half of those workers retain the previous local state until fresh reseeding resumes. This is not a caricatured broken RNG; it is a deployment-style fault model aimed at image reuse and brownout recovery.

The protocol has two parts. First, we validate the law $P(\eta)$ with Monte Carlo over $\eta\in\{0.75,0.80,\ldots,1.00\}$. Second, we run the scheduler simulation and estimate $\hat{\eta}$ from unique work-unit identifiers and $\hat{\rho}$ from post-boot token diversity. To make replication explicit, the package accompanying this submission includes the figure file used in the paper and a compact Python script that regenerates the curve and the summary metrics from the parameter set in Table~\ref{tab:protocol}.

\begin{table}[t]
\caption{Simulation-derived scheduler metrics with 95\% confidence intervals.}
\centering
\footnotesize
\setlength{\tabcolsep}{4pt}
\begin{tabular}{>{\raggedright\arraybackslash}p{0.30\linewidth}
                >{\centering\arraybackslash}p{0.20\linewidth}
                >{\centering\arraybackslash}p{0.20\linewidth}
                >{\centering\arraybackslash}p{0.18\linewidth}}
\toprule
Baseline & $\hat{\eta}$ & $\hat{\rho}$ & $\hat{P}(\hat{\eta})/P(1)$ \\
\midrule
Deterministic (B1)      & 1.000            & 1.000            & 1.000 \\
Strong classical (B2)   & 1.000            & 1.000            & 1.000 \\
QRNG-assisted (B3)      & 1.000            & 1.000            & 1.000 \\
Correlated-restart (B4) & $0.962\pm0.0004$ & $0.899\pm0.0006$ & 0.978 \\
\bottomrule
\end{tabular}
\label{tab:results}
\end{table}

\begin{figure}[t]
\centering
\includegraphics[width=\linewidth]{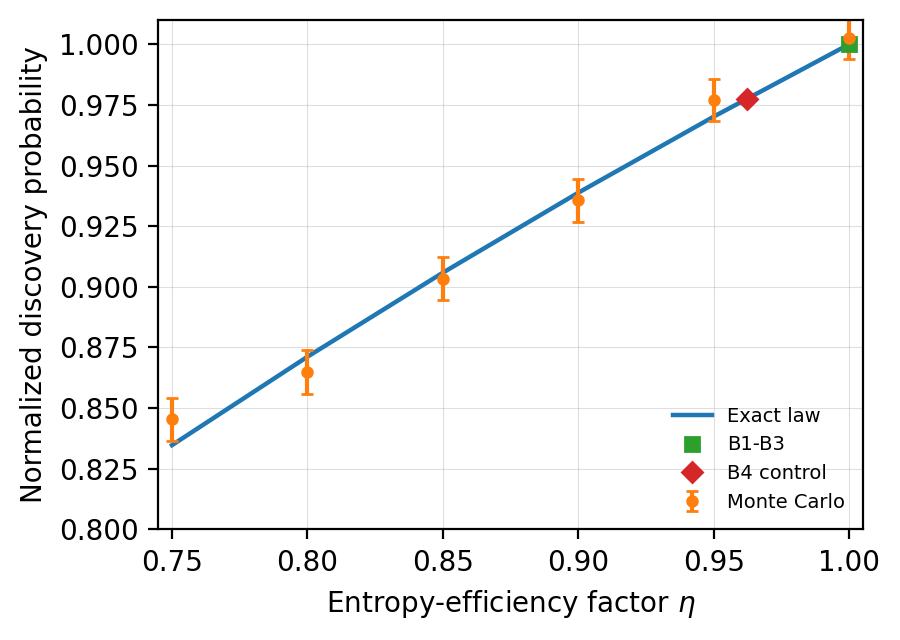}
\caption{Normalized discovery probability versus entropy-efficiency factor $\eta$. Monte Carlo points with 95\% confidence intervals follow the exact law. Competent baselines B1--B3 cluster at $\eta\approx 1$, while the correlated-restart control B4 falls on the same curve at lower effective coverage.}
\label{fig:eta}
\end{figure}

Figure~\ref{fig:eta} confirms the analytical law and separates two questions that are often conflated: whether good entropy engineering prevents avoidable overlap, and whether QRNG changes the PoW law itself. The answer to the first is yes under failure-prone infrastructure; the answer to the second is no. Table~\ref{tab:results} makes the main simulation result precise: under competent design, deterministic, strong-classical, and modeled QRNG-assisted baselines are indistinguishable because all keep $\eta\approx 1$. By contrast, the correlated-restart control shows measurable losses in both $\eta$ and $\rho$, while the normalized success rate remains on the exact curve predicted by $P(\eta)$.

A sensitivity sweep on the correlated-restart fraction from 0 to 30\% moved $\hat{\eta}$ from 1.000 to 0.928 and $\hat{\rho}$ from 1.000 to 0.852, with the measured normalized success probability staying within 0.003 of the exact curve throughout. This graded response strengthens the evidential section beyond a single stress point and shows that the observables track the theory monotonically.

\section{Discussion}

The results do not imply that QRNGs are irrelevant to mining ecosystems. Rather, they show that the appropriate benchmark is \emph{assurance}, not consensus acceleration. QRNGs can strengthen the trustworthiness of the entropy root used to instantiate or reseed deterministic services, improve restart robustness in distributed controller fleets, and support defensible claims about entropy provenance in audited devices \cite{sp90b,sp90c,huang,blanco2025,blanco2026}. These are legitimate and technically meaningful roles for a quantum primitive within hybrid quantum-classical infrastructure.

This distinction is methodologically important. A rigorous contribution in quantum technologies does not need to claim a direct algorithmic advantage over a mature classical workload. It can instead identify a technically justified insertion point for a quantum primitive, define observables for future validation, and compare that primitive fairly against competent classical baselines. In that sense, the present work is closer to quantum information engineering, secure quantum-classical integration, and deployable infrastructure design than to a generic blockchain optimization study.

The threat model is intentionally narrow. We assume an honest PoW engine and focus on failures in randomness-dependent support services, including repeated post-boot state, predictable namespace seeds, image-based recovery, and avoidable overlap. We do not model selfish mining, block withholding, ASIC reverse engineering, or device-level latency and cost profiles \cite{eyal,minersdilemma,luu,devries}. These limitations define the scope of the analysis rather than weaken its central result.

Trusted entropy throughput should not be confused with PoW throughput. Let $R_h$ denote the nominal header-testing rate of the mining engine, let $s$ be the number of entropy bits injected into a DRBG or namespace generator at each reseed, and let $\tau$ be the reseed interval. The required trusted entropy throughput is
\begin{equation}
R_{\mathrm{req}} = \frac{s}{\tau}.
\end{equation}
In realistic deployments, $R_{\mathrm{req}}$ can be orders of magnitude smaller than $R_h$. This clarifies why QRNG need not operate inside the per-hash loop to be useful: it can serve at the entropy root of the control plane while deterministic logic continues to drive the high-rate PoW engine.

\begin{table}[t]
\caption{Deployment-oriented interpretation of QRNG adoption.}
\centering
\footnotesize
\begin{tabular}{>{\raggedright\arraybackslash}p{0.36\linewidth} p{0.54\linewidth}}
\toprule
Scenario & Recommended interpretation \\
\midrule
Single well-partitioned miner & No throughput case for QRNG; adoption is mainly justified when assurance or certification requirements dominate. \\
Coordinated pool or controller fleet & Strong entropy helps preserve namespace freshness, identifier uniqueness, and restart diversity. \\
Audited appliance in a mining stack & QRNG is justified by entropy provenance, defense-in-depth, and stronger trust in the entropy root. \\
\bottomrule
\end{tabular}
\label{tab:deploy}
\end{table}

The main limitation of the present study is the absence of device-level validation using physical QRNG hardware or recorded QRNG bitstreams. The QRNG-assisted baseline should therefore be interpreted as an idealized entropy-root model, not as a characterization of a specific QRNG implementation. A natural next step is a device-in-the-loop replication in which the scheduler is reseeded using raw or conditioned output from an actual QRNG. Such a study should record the QRNG source model, raw-bit throughput, conditioning method, health-test results, estimated min-entropy, reseed cadence, latency distribution, and failure/restart behavior. The resulting operational logs would allow direct comparison of $\eta$, $\rho$, and $R_{\mathrm{req}}$ against the deterministic and strong-classical baselines defined here.

A final methodological lesson is that fair benchmarking of quantum entropy in hybrid infrastructure requires strict control of confounders. Scheduler logic, work-unit format, nominal work volume, namespace size, and reseed cadence must remain fixed while only the entropy root varies; otherwise, a QRNG design may appear to outperform alternatives simply because it was coupled to a better scheduler or a broader namespace policy. Reported results should therefore include $\eta$ and $\rho$ alongside discovery probability, so that any measured gain can be attributed to effective coverage or post-boot diversity rather than being misread as a change in PoW itself.

The benchmarking logic developed here extends beyond the specific Bitcoin case studied in this paper. The same issue arises whenever a quantum primitive is inserted into a larger deterministic workload after initialization, including QRNG-backed entropy services, embedded post-quantum stacks, and controller layers for quantum communication systems. In such settings, the central question is rarely whether the quantum primitive accelerates the main computation. The relevant question is whether it improves the quality, traceability, and resilience of the randomness-dependent services that surround that computation.

\section{Conclusion}

For honest Bitcoin PoW, QRNG substitution is consensus-neutral: the block-discovery law depends on the number of distinct tested headers, not on whether candidate ordering is deterministic, classically pseudorandom, or QRNG-assisted. The contribution of this paper is therefore methodological, operational, and reproducible rather than protocol-altering or device-level. By introducing $\eta$ and $\rho$, benchmarking QRNG against a strong classical baseline under identical scheduler logic, and incorporating a realistic correlated-restart control with confidence-bounded simulations, we show where quantum entropy can matter and how its effect can be measured. The practical implication is direct: QRNGs are most strongly justified as high-assurance entropy roots for mining-adjacent hybrid infrastructure, particularly in seeding, provisioning, restart robustness, and auditable entropy provenance.

From a deployment perspective, the decision rule is straightforward. If a mining stack already partitions namespaces deterministically, recovers cleanly from reboots, and does not require strong external assurance arguments, then a well-engineered classical entropy source combined with a validated DRBG may be sufficient. If, however, operators require auditable entropy provenance, robust synchronized-restart recovery, or defensible reseeding policy across distributed controllers, then QRNG becomes technically justified without any need to invoke a fictitious consensus accelerator. In this sense, the negative result on PoW speedup becomes a positive engineering guideline for real deployments.

More broadly, the methodological lesson extends beyond the specific case studied here. Many near-term quantum technologies are likely to appear first as carefully delimited subsystems embedded within larger classical infrastructures, rather than as stand-alone replacements for entire computational workloads. A rigorous contribution in such settings is one that identifies the appropriate insertion point, states clearly what does and does not improve, and defines observables that enable independent validation. The present work argues that QRNG in Bitcoin-related infrastructure should be evaluated precisely in those terms.

Future work should therefore prioritize hardware-in-the-loop validation with actual QRNG devices or recorded QRNG streams, allowing the proposed observables to be tested under real entropy throughput, latency, conditioning, and health-test constraints.

\end{document}